\documentclass[prc,aps,twocolumn,showpacs]{revtex4}
\usepackage{epsf}
\newcommand{\ds}{\displaystyle}
\newcommand{\dsf}{\ds\frac}

\newcommand{\re}[1]{(\ref{#1})}

\begin{document}
\sloppy

\title{Meson-Nucleon Vertex Form Factors
at Finite Temperature Using a Soft Pion Form Factor}

\author{U.T. Yakhshiev$^{a}$ A.W. Thomas$^{b}$, and F.C. Khanna$^{b,c}$}

\affiliation{${}^{a}$Theoretical Physics Department and
Institute of Applied Physics,\\
 National University of Uzbekistan,
Tashkent-174, Uzbekistan,\\
${}^{b}$Special Research Centre for the Subatomic Structure of Matter,
\\
University of Adelaide, South Australia 5005, Australia\\
${}^{c}$Physics Department, University of Alberta,
Edmonton, Alberta, Canada, T6G 2J1,\\
and \\
TRIUMF, 4004 Wesbrook Mall, Vancouver, British Columbia, Canada, V6T 2A3}

\date{\today}

\begin{abstract}
The temperature and density dependence of the meson-nucleon vertex form
factors is studied in the framework of thermofield dynamics.
Results are obtained for two rather different nucleon-nucleon
potentials:
the usual Bonn potential and the variation
with a softer $\pi NN$ form factor, due to
Holinde and Thomas. In general, the results show only a modest
degree of sensitivity to the choice of interaction.
\end{abstract}


\maketitle


Recently~\cite{1}
the temperature dependence of the meson nucleon form factors
has been calculated in Thermofield Dynamics (TFD) using the Bonn $N-N$
interaction~\cite{2} that includes exchanges of $\pi, \sigma, \rho$ and
$\omega$ mesons, with
the $\pi NN$ vertex form factor having a monopole form
\begin{equation}
G_{\pi NN}(q^{2}) = g_{\pi NN}
\frac{\Lambda^{2} - m^{2}_{\pi}}{\Lambda^{2} + q^{2}} ,
\end{equation}
%
where $\Lambda = 1.3$ GeV for the $\pi NN$ vertex.
It is anticipated that in low energy $N-N$ scattering $q^{0} \approx 0$ so that
the 4-momentum transfer squared is usually minus the 3-momentum
transfer squared. However it has been argued~\cite{3}
that there are several
{}factors that point to a significantly softer form factor with
$\Lambda \approx 700\div 800$~MeV. It was concluded that the $\pi NN$
{}form factor for a free nucleon is close to the measured axial-vector
{}form factor, thus giving $ \Lambda \approx 500\div
800$ MeV. It has been suggested that a soft $\pi NN$
{}form factor would make it difficult to fit the $N-N$ scattering and,
in particular the properties of the deuteron. However Holinde and
Thomas~\cite{3,4} established that a good fit can be obtained for
$\Lambda = 800$~MeV.

The study of the temperature and density dependence of the coupling
constants, form factors and the critical temperature, where the
coupling constant goes to zero, is needed for this softer $\pi NN$
form factor. This would clarify any dependence on the choice of the
interaction and give us a better understanding of the
equation of state that may be relevant for heavy-ion collisions. The
behaviour of nuclear matter at finite temperature and density and
the phase transition from the hadronic to the quark-gluon phase with
subsequent hadronisation can provide valuable information about the
nature of confinement in QCD.

In this brief communication we present results for the temperature and
density dependence of the meson-nucleon couplings,
using the OBEPTI $N-N$ interaction~\cite{3,4},
and compare them to the results obtained earlier with the BONN potential,
OBEP~\cite{2}.
The main calculational procedure has been described
earlier~\cite{1} and we concentrate on discussing
the results in the present note.

\begin{figure*}[htb]
\vspace{-1.cm}
\begin{center}
   \epsfxsize=15cm
\epsffile{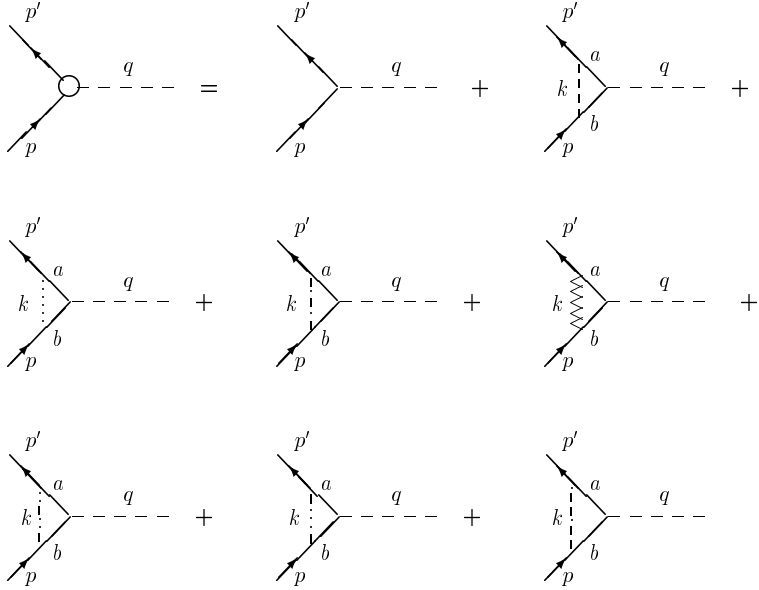}
\end{center}
   \vspace{-12.cm}
   \caption{
   \label{fig1}
   Feynman diagrams for the pion-nucleon vertex.
   The solid line indicates the nucleon, while the dashed,
   dotted, dot-dashed, wavy, dot-dot-dashed, dot-dot-dot-dashed and
   dashed-dashed-dot lines indicate the $\pi-$,
   $\rho-$, $\omega-$, $\sigma-$, $\pi'-$,
   $\eta-$ and $\delta-$ mesons, respectively.}
\end{figure*}

The temperature dependence is calculated within Thermofield Dynamics,
a finite temperature field theory. The finite temperature modifications
of the vertex
functions are calculated from the one-meson-exchange Feynmann diagrams
shown in Fig.~\ref{fig1}.
The mathematical representation was explained earlier~\cite{1}. To
initiate the calculations, zero temperature coupling constants and
vertex form factors are chosen to be either
i)the Bonn potential, OBEP, or
ii) the modified Bonn
potential with a soft $\pi NN$ form factor, OBEPTI, due to Holinde and
Thomas (see Table~\ref{potpar}).
\begin{table}
\begin{center}
\begin{ruledtabular}
\begin{tabular}{cccccc}
Meson & $I(J^P)$ & Mass (GeV) & $\Lambda_\alpha$ (GeV) &
$g_\alpha^2/4\pi$ & $f_\alpha/g_\alpha$\\
\hline
\multicolumn{6}{c}{OBEP parametrization}\\
\cline{2-6}
$\pi$  & $1(0^-)$ & 0.138 & 1.05  & 14.9 & $\dots$\\
$\rho$ & $1(1^-)$ & 0.769 & 1.3   & 0.99& 6.1\\
$\omega$ & $0(1^-)$ & 0.783 & 1.5 & 20.0 & $\dots$\\
$\sigma$ & $0(0^+)$ & 0.550 & 2.0 & 8.383 & $\dots$\\
\cline{2-6}
\multicolumn{6}{c}{OBEPTI parametrization}\\
\cline{2-6}
$\pi$  & $1(0^-)$ & 0.138 & 0.8 & 14.6 & $\dots$\\
$\pi'$ & $1(0^-)$ & 1.200 & 2.0 & 100 & $\dots$\\
$\eta$ & $0(0^-)$ & 0.549 & 1.5 & 5.0 & $\dots$\\
$\rho$ & $1(1^-)$ & 0.769 & 1.3 & 0.92& 6.6\\
$\omega$ & $0(1^-)$ & 0.783 & 1.5 & 20.0 & $\dots$\\
$\delta$ & $1(0^+)$ & 0.983 & 2.0 & 2.881 & $\dots$\\
$\sigma$ & $0(0^+)$ & 0.550 & 2.0 & 8.383 & $\dots$\\
\end{tabular}
\end{ruledtabular}
\end{center}
\caption{
\label{potpar}
Parameters for the Bonn potential OBEP~[2]
and for the modified Bonn potential OBEPTI~[4].
}
\end{table}
It is important to note that results for $\pi^{\prime}$, appearing in
OBEPTI, can be obtained from those of the $\pi$-meson by the
following procedure
\begin{equation}
\begin{array}{lll}
f(\vec{q}^{\,2}) &=&
\dsf{G_{\pi^\prime NN}(\vec{q}^{\,2},T,p)}
{G_{\pi NN}(\vec{q}^{\,2}, T, p)} =
\dsf{G_{\pi^\prime NN}(\vec{q}^{\,2}, 0, 0)}
{G_{\pi NN}(\vec{q}^{\,2}, 0, 0)}
\\\quad\\
 &=&
\dsf{g_{\pi^\prime NN}}{g_{\pi NN}}
\left(\dsf{\Lambda^{2}_{\pi^\prime} -m^{2}_{\pi^\prime}}
{\Lambda^{2}_{\pi} -
m^{2}_{\pi}}\right) \left(\frac{\Lambda^{2}_{\pi} +
\vec{q}^{\,2}}{\Lambda^{2}_{\pi \prime} + \vec{q}^{\,2}}\right)\,.
\end{array}
\label{scalingfac}
\end{equation}

\begin{figure*}[htb]
\vspace{-5cm}

\begin{center}
   \epsfxsize=15cm
\epsffile{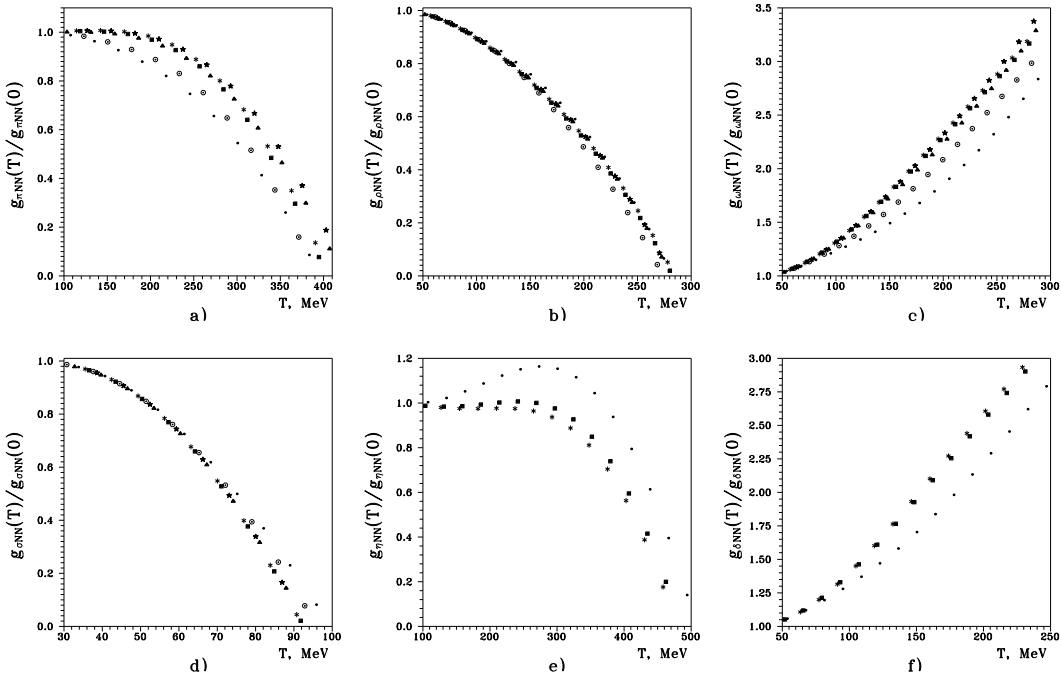}
\end{center}

   \vspace{-7cm}
   \caption{
   \label{fig2}
   The ratio of meson-nucleon
   coupling constants at finite temperature, $T$, to those at $T=0$, as a
   function of temperature at $\rho=0$ (stars and fivestars),
   $\rho=\rho_0$ (squares and triangles), $\rho=5\rho_0$
   (dotted circles and simple dots). Stars, squares and dotted
   circles correspond to calculations using the OBEP
   potential~\protect\cite{2}.
   {}Fivestars, triangles and dots correspond
   to calculations using the OBEPTI potential~\protect\cite{4}.
   Results for the $\pi'$ coincide with the OBEPTI part of figure {\bf a}
   (see Eq.~\re{scalingfac}).}
\end{figure*}

The ratios of the coupling constants, $g_{BNN}(T)/g_{BNN}(0)$, are plotted
in Fig.~\ref{fig2}.
Note that the $\eta , \delta$ and $\pi^\prime$ mesons are present only for
the potential of Holinde and Thomas. We observe
that for the $\pi -, \pi^\prime -, \rho -, \sigma-$ and
$\eta$-mesons this ratio goes through zero
at different temperature depending on
the density. This temperature is called the critical temperature,
$T^{B}_{C}$. For $\omega-$ and $\delta-$ mesons the behaviour of the ratio
is opposite -- i.e. it increases instead of going through zero. For
$\pi-$ mesons the difference
between the two potentials is smallest at a density of order $\sim 5
\rho_{0}$, where $\rho_0$ is normal nuclear matter density.

\begin{figure*}[htb]
\vspace{-5cm}

\begin{center}
   \epsfxsize=15cm
\epsffile{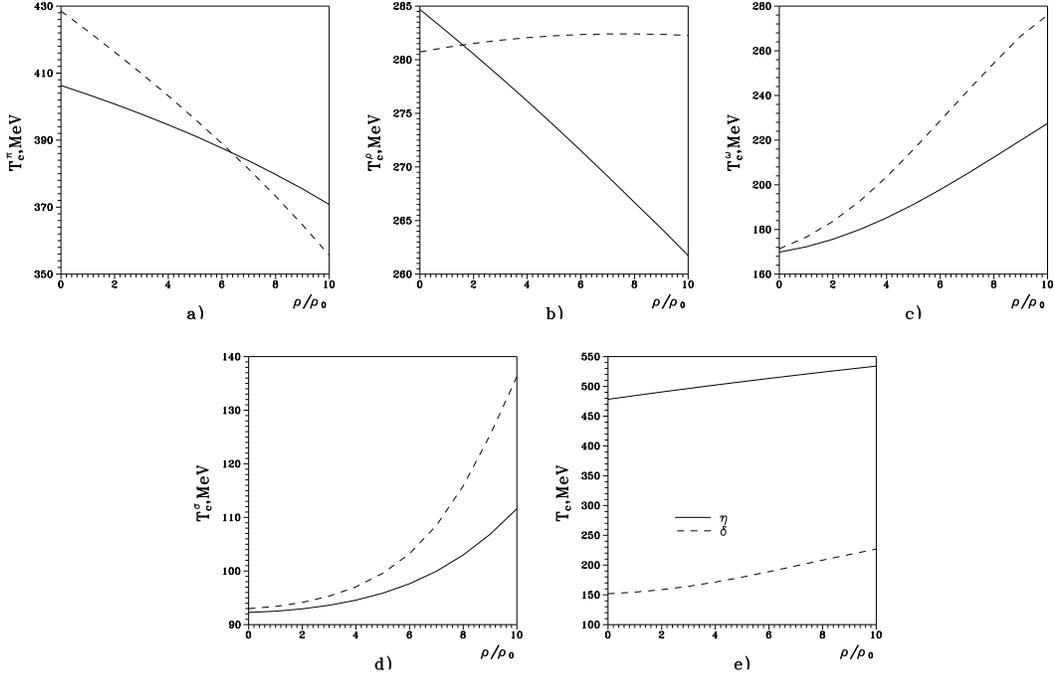}
\end{center}

   \vspace{-7.5cm}
   \caption{
\label{fig3}
   Density dependence
   of the critical temperature for several ({\bf a} - $\pi$,
   {\bf b} - $\rho$, {\bf c} - $\omega$, {\bf d} - $\sigma$,
   {\bf e} - $\eta,\delta$) mesons.
   In figures {\bf a - d} the solid lines correspond to the calculation
   based on the OBEP potential~\protect\cite{2}, while the dashed lines
   correspond
   to calculations using the OBEPTI potential~\protect\cite{4}.
   In figure {\bf e} the solid line corresponds to the $\eta$ and the dashed
   line to the $\delta$. Calculations
   in figure {\bf e} are made using the OBEPTI potential. The results
   for the $\pi'$ coincide with the OBEPTI part of figure {\bf a}.}
\end{figure*}

The changes with temperature for the two sets of
parameters are usually quite
similar, though the actual numbers differ. In
Fig.~\ref{fig3} we show
the critical temperatures as a function of the density of  nuclear
matter. The behaviour of $T^{B}_{c}$ for various mesons
using the two potentials is quite different. For the $\pi-$ and $\sigma-$
mesons, $T^{B}_{c}$ changes more rapidly for the softer pion vertex, while
for the $\rho-$meson $T^{B}_{c}$ is practically constant -- in
contrast with the OBEP potential, where it changes rapidly. The changes
in $T^{\omega}_{c}$ show
quite a significant dependence on the choice of potential.

\begin{figure*}[htb]
\vspace{-5cm}

\begin{center}
   \epsfxsize=15cm
\epsffile{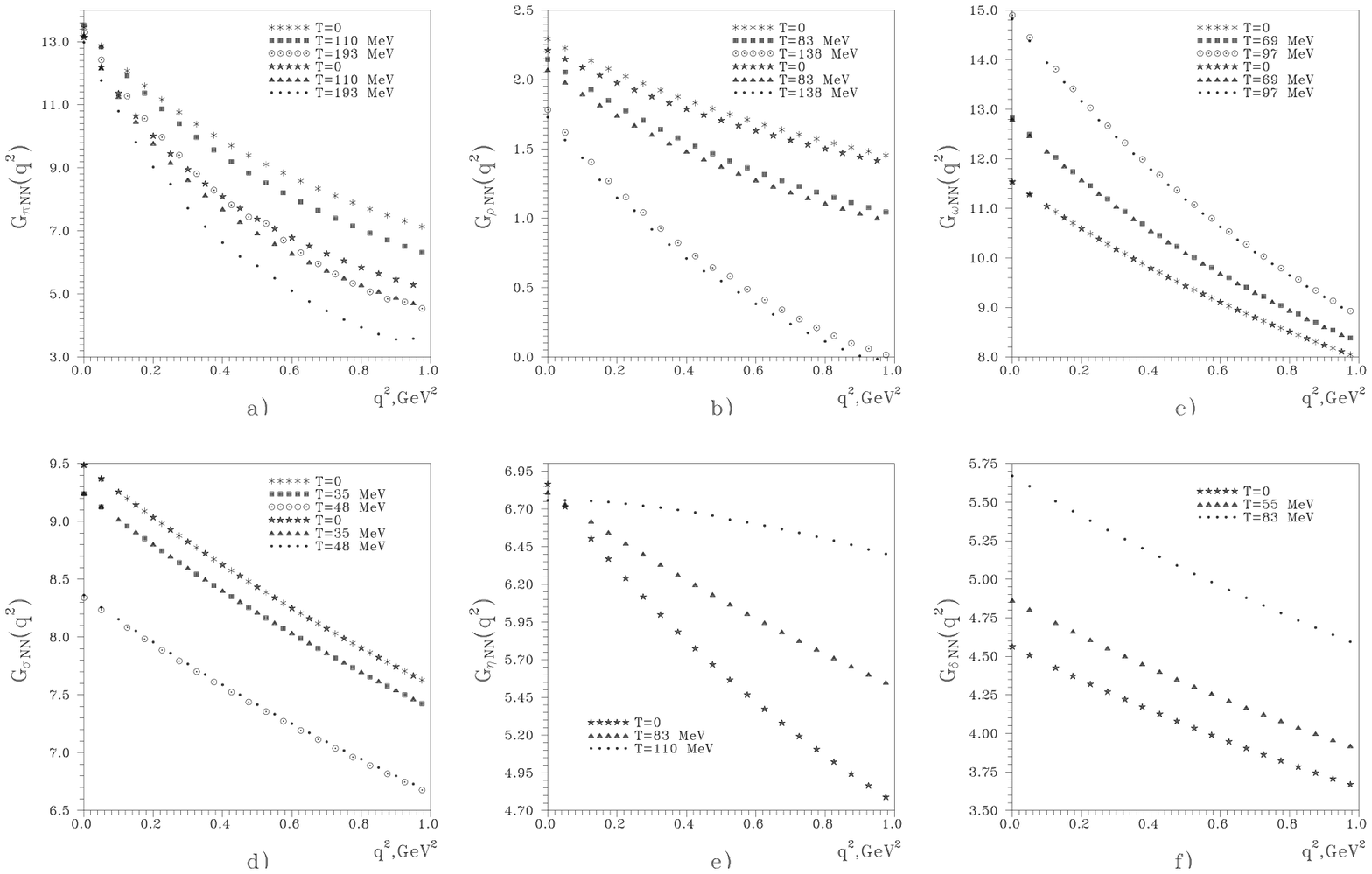}
\end{center}

   \vspace{-7.5cm}
   \caption{
\label{fig4}
   Meson-nucleon form factors at several temperatures as a function of $q^2$
   using OBEP~\protect\cite{2}
   (stars, squares and dotted circles) and OBEPTI
   potentials~\protect\cite{4}
   (fivestars, triangles and dots) at $\rho=0$.
   Results for the $\pi'$ can be extracted
   from the OBEPTI part of figure {\bf a}, taking into account the scaling
   function, $f(q^2)$ (see Eq.~\re{scalingfac}).}
\end{figure*}

We have also investigated the $q^{2}$-dependence of the
meson-nucleon form factors with temperature. The results
of these calculations presented in Fig.~\ref{fig4}.
In this case the pion form
factor shows a large change and a stronger dependence on the choice
of $T=0$ parameter. For other mesons the form-factors show very
little dependence on a hard or soft pion vertex function.

It is clear that at
density $\rho_{0}$, the behaviour of the critical temperature for the
$\rho$ meson is dramatically changed when the parameters of the OBE potential
are changed. In other words, $g_{\rho NN}$ is sensitive to
parameter changes. For the OBEPTI potential $T^{\rho}_{c}(\rho)$ is
almost constant, while for the OBEP potential $T^{\rho}_{c}(\rho)$
decreases rapidly. For the $\pi$ meson $T_{c}(\rho)$ decreases even
more rapidly with increasing density. For other mesons
$T_{c}(\rho)$ has the same
behaviour, increasing with increasing density. For all mesons
except the pion, the critical temperature increases when one uses
the OBEPTI parametrization. Nevertheless, at moderate densities the critical
temperature for the $\pi$ is also somewhat higher.

In Fig.~\ref{fig4} 
we present the form factors at several temperatures at zero
density. The form factors of the $\pi NN$ and $\rho NN$ vertices are more
sensitive to parameter changes, while $G_{\omega NN}$ and $G_{\sigma
NN}$ are not. This is also seen from Fig.~\ref{fig3},
where the critical
temperature for the $\omega$ and $\sigma$ mesons is almost the same for both
parameter sets at zero density. At high densities
$G_{\pi NN}$, and $G_{\rho NN}$ fall off more rapidly at high momentum
transfer, constituting quenching of the cut off parameters for
corresponding meson form factors.

To summarise our results we present a practical parametrization of the
temperature and density dependence of the form factors.
This may be useful for other applications. At small momentum
transfers we can parametrize them by a monopole form
\begin{equation}
G_{BNN} (\vec{q}^{\,2}, T, \rho ) = g_{B}(T, \rho)
\frac{\Lambda^{2}_{B} (T, \rho) - m^{2}_{B}}{\Lambda^{2}_{B}(T,
\rho) + \vec{q}^{\,2}}\,,
\label{parff}
\end{equation}
%
where we still use the mass of the
corresponding meson at zero temperature and density .
In general the effective mass of a meson, $m_{B}$, is
also temperature dependent, but that has not been considered here. From
Fig.~\ref{fig3} we see that $T_{c}$ has an almost
linear dependence on density, except
for the case of $\sigma$-meson. However,
even for this case, at moderate densities
there is an almost linear dependence. Consequently, we can write the following
density dependence of the critical temperature
\begin{equation}
T_{c} = T_{0} \left(1 + D_{B} \dsf{\rho}{\rho_{0}}\right)\,.
\label{Tc}
\end{equation}
{}Furthermore, using a polynomial form for the ratios
\begin{eqnarray}
\label{par}
\frac{g_{B}(T, \rho)}{g_{B}(T = 0, \rho = 0)} = \frac{1 + (T/T_{c})
\alpha^{g}_{B} + (T/T_{c})^{2} \beta^{g}_{B}}{1 + (\rho /\rho_{0})
C^{g}_{B}}, \nonumber \\
\frac{\Lambda_{B}(T, \rho)}{\Lambda_{B}(T = 0, \rho = 0)} = \frac{1 +
(T/T_{C}) \alpha^{\Lambda}_{B} + (T/T_{c})^{2}
\beta^{\Lambda}_{B}}{1 + (\rho / \rho_{0})C^{\Lambda}_{B}}
\end{eqnarray}
we find the values of the parameters given in Table~\ref{paramset}.
Again it is clear that $G_{\pi \prime NN}$ can be restored from $G_{\pi
NN}$ as given in Eq.~(2).
\begin{table}
\begin{center}
\begin{ruledtabular}
\begin{tabular}{ccccccccc}
Meson &
$\alpha^g$ & $\beta^g$ & $C^g$ &
$\alpha^\Lambda$ & $\beta^\Lambda$ & $C^\Lambda$ &
$D$ & $T_c$ \\
\hline
\multicolumn{9}{c}{OBEP parametrization}\\
\cline{2-9}
$\pi$    & 1.329 &-2.390 & 0.033 &-0.230 &-0.400 &-0.008 &-0.008 & 406\\
$\rho$   & 0.066 &-1.058 &-0.002 &-0.070 &-1.976 &-0.007 &-0.008 & 285\\
$\omega$ & 0.138 & 0.808 &-0.008 & 0.008 &-0.524 & 0.006 & 0.031 & 170\\
$\sigma$ & 0.559 &-1.547 & 0.010 &-0.018 &-0.024 &-0.003 & 0.005 &  93\\
\cline{2-9}
\multicolumn{9}{c}{OBEPTI parametrization}\\
\cline{2-9}
$\pi$    & 1.018 &-2.029 & 0.031 &-0.142 &-0.369 &-0.007 &-0.016 & 429\\
$\rho$   & 0.108 &-1.079 &-0.001 &-0.081 &-2.012 &-0.006 & 0.001 & 281\\
$\omega$ & 0.062 & 0.904 &-0.013 & 0.005 &-0.547 & 0.008 & 0.058 & 171\\
$\sigma$ & 0.554 &-1.528 & 0.016 &-0.008 &-0.043 &-0.002 & 0.007 &  93\\
$\eta$   & 1.449 &-2.442 &-0.020 &-0.963 &14.468 &-0.003 & 0.012 & 478\\
$\delta$ & 0.211 & 0.724 &-0.014 & 0.233 &-0.364 & 0.019 & 0.045 & 152\\
\end{tabular}
\end{ruledtabular}
\end{center}
\caption{Parameters of vertex form factors in Eqs.~(4)-(5).
The last column is the critical temperature, $T_c$, at zero density,
$\rho=0$.}
\label{paramset}
\end{table}

In conclusion, the choice of hard and soft $\pi NN$ form factors does
change the behaviour of the critical temperature with density. While the
changes are not, in general, dramatic, there is a
qualitative change for the $\rho$ meson. When it comes to the
$q^{2}$-dependence
of the form-factors, the changes for all mesons are similar.
The parametrization of the changes with temperature and density of the
boson coupling
constants to the nucleon, as well as the cut-off
parameters in the form factors,
provides a useful way to summarise the
variation of these parameters. Finally it is important to realize
that while the couplings of the mesons to nucleons
decrease to zero at a critical
temperature, it is possible that before we reach that temperature,
the system may have already undergone a phase transition,
leading to a deconfined quark-gluon state.

\section*{Acknowledgments}
U.T. Yakhshiev acknowledge International Centre for Theoretical
Physics (ICTP) for the hospitality during his stay at the Centre.
The work of U.T. Yakhshiev was supported in part by INTAS (YSF
00-51). F.C. Khanna would like to thank Tony Thomas for the
hospitality at CSSM where the main ideas were considered. The research
of F.C.K. is partly supported by NSERC. Finally, this work was supported
in part by the Australian Research Council.



\begin{thebibliography}{99}
\bibitem{1}
   A.M.\,Rakhimov,  U.T.\,Yakhshiev, F.C.\,Khanna.
        Phys. Rev. {\bf C61}, 024907 (2000).
\bibitem{2}
   R.\,Machleidt,
        Adv. Nucl. Phys. {\bf 19}, 189 (1989).
\bibitem{3}
   A.W.\,Thomas, K.\,Holinde.
        Phys. Rev. Lett. {\bf 63}, 2025 (1989);
P.~A.~Guichon, G.~A.~Miller and A.~W.~Thomas,
Phys.\ Lett.\ B {\bf 124}, 109 (1983).
\bibitem{4}
   K.\,Holinde, A.W.\,Thomas,
        Phys. Rev. {\bf C42}, R1195 (1990).
\end{thebibliography}
\end{document}